\documentclass[review]{elsarticle}

\usepackage{amsmath}
\usepackage{amsbsy}
\usepackage{amssymb}
\usepackage{color}
\usepackage{epsfig}
\usepackage{setspace}
\usepackage{bm}
\usepackage{natbib}

\newcommand{\prob}{P}
\newcommand{\mbf}{\mathbf}
\newcommand{\bs}{\boldsymbol}
\newcommand{\E}{I\!\!E\hspace{.03cm}}

\journal{ArXiv}

\begin{document}

\begin{frontmatter}

\title{Penalised maximum likelihood estimation in multistate models for interval-censored data}

\author{Robson J. M. Machado} 
\ead{robsonjmachado@gmail.com}
\author{Ardo van den Hout }
\author{Giampiero Marra }

\address{Department of Statistical Science, University College London, Gower Street, \\ London WC1E 6BT, UK.} 





\begin{abstract}
Multistate models can be used to describe transitions over time across states. In the presence of interval-censored times for transitions, the likelihood is constructed using transition probabilities. Models are specified using proportional hazards model for the transitions. Time-dependency is usually defined by parametric models, which can be too restrictive. Nonparametric hazards specification with splines allow for flexible modelling of time-dependency without making strong model assumptions. Penalised maximum likelihood is used to estimate the models. Selecting the optimal amount of smoothing is challenging as the problem involves multiple penalties. We propose an automatic and efficient method to estimate multistate models with splines in the presence of interval-censoring. The method is illustrated with a data analysis and a simulation study.
\end{abstract}

\begin{keyword}
 Automatic smoothing parameters estimation\sep Interval-censoring \sep Markov models \sep Panel data \sep Splines.
\end{keyword}

\end{frontmatter}


\section{Introduction}\label{S:1}
\noindent In biostatistics, disease progression can be described using patient's health status over time. Multistate models are commonly used to describe transitions across a set of discrete states.  Time of transitions are usually observed intermittently leading to interval-cesored data. To facilitate estimation, a time homogeneous Markov process is usually assumed \citep{kalbfleisch1985, jackson2011multi}. For a wide range of applications, the risks of moving across states depend on the current state and on time. In this case, a non-homogeneous Markov assumption is assumed to model the multistate process. Several time-dependent models can be fitted with parametric specifications \cite{van2014multi}. However, the functional form underlying the data is often unknown and parametric models can be too restrictive.   

A penalised maximum likelihood estimation for a progressive three-state model is developed in Joly and Commenges \cite{joly1999penalized}. Estimation is performed with an algorithm which uses analytical derivatives of the penalised log-likelihood. The smoothing parameters are selected using a grid search with cross-validation. In this case, models have to be fitted for every combination of smoothing parameters defined by the grid. Joly et al. \cite{joly2002penalized} use the same approach for an illness-death model. The method used in both works can be computationally extensive for models with multiple smoothing parameters. In addition, the method requires explicit expressions for the probabilities transitions. Calculating those formulae can be intractable for more complex models, such as models with more than four states and recovery  \citep{jackson2011multi}. Titman \cite{titman2011} uses a numerical approximation to calculate the transition probabilities at the level of the corresponding differential equations. The method allows for nonparametric hazard specifications with $B$-splines placed at equidistant knots. However, the log-likelihood is maximised without penalisation. Machado and Van den Hout \cite{machado2017flexible} proposed a penalised likelihood method to estimate semiparametric multistate models with $P$-splines. The smoothing parameters are selected by using grid search. Even though the method is general and allows for backward transitions, it can become burdensome for applications that involve multiple penalties.

 
In this paper, we propose an efficient method to estimate nonparametric multistate models with splines for interval-censored data.  A Markov process framework is used to formulate the models. Hazards are specified with splines base functions to allow for flexible modelling over time. Estimation is undertaken using a penalised likelihood approach. Given a piecewise-constant approximation to the hazards, the Fisher scoring algorithm presented in Kalbfleisch and Lawless \cite{kalbfleisch1985} can be adapted for time-dependent models. An automatic method is used to estimate the multiple smoothing parameters. The new estimation procedure is made possible by rewriting the optimisation problem in a generalised likelihood based penalised method \cite{marra2016simultaneous}. The fitted multistate model with splines can be used for flexible modelling of time-dependency, but also to check parametric specifications. 
\begin{figure}[t!]\centering
\includegraphics[scale = 1 ]{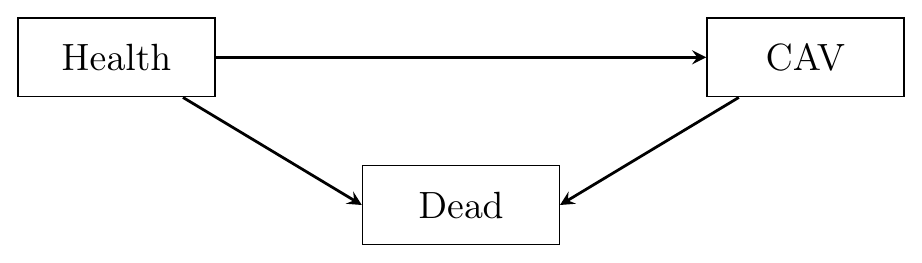}
\caption{\label{fig1} Three-state model for disease progression after transplant}
\end{figure}



\subsection{Cardiac allograft vasculopathy (CAV) data}\label{cav_intro}
\noindent The method is applied to data for cardiac allograft vasculopathy (CAV). The data come from Papworth Hospital U.K. and are available in the \textsf{msm} package \cite{jackson2011multi}. CAV is a narrowing of the arterial walls and the main cause of death in heart transplantation patients. The data are a series of approximately yearly angiographic examinations of heart transplant recipients. The state at each time is a grade of CAV which can be normal, moderate or severe. Dead is the absorbing state and time of death is known within one day. The data contain 2816 rows which are grouped by 614 patients and ordered by years after transplant. Each row represents an examination and contains additional covariates. The process is biologically irreversible and of particular interest is the onset of CAV. Diagnosis of ischaemic heart disease (IHD) and donor age are known to be major risk factors of disease onset \cite{titman2011}. 
In order to investigate this, three-state progressive models can be defined. The states are classified as normal (1) if the patient has not developed the disease, ill (2) if the patient has developed moderate or severe CAV and dead (3) if the patient has died, see Figure \ref{fig1}. Follow-up data after 15 years are not used, since after this time data are scarce which may cause identifiability problems. Titman \cite{titman2011} used a similar formatting of the CAV data.  Table \ref{statetable} gives the number of times each pair of states was observed at successive observation times.
\begin{table}
\caption{State table for the CAV data: number of times each pair of states was observed at successive observation times. The two living states are defined by CAV severity}
\begin{center}
\begin{tabular}[H]{lccr}
\hline
 &  \multicolumn{3}{l}{To } \\
From        & 1& 2 & 3  \\
\hline
 1  &  1314 & 223 & 136 \\
  2 & 0 & 411 & 105 \\ \hline

\end{tabular} \label{statetable}
\end{center}

\end{table}

\section{Multistate models with splines}\label{sec:models}
\subsection{Model representation}
\noindent Let $Y(t)$ be a continuous-time Markov chain on finite state space $\mathcal{S}$, time-homogeneous transition probabilities
are given by \[
p_{rs}(t)=P\big(Y(t + u)=s|Y(u)=r\big),
\]
for $r,s \in \mathcal{S}$, $u\geq 0$ and $t\geq 0$. This Markov chain is time-homogeneous because the probability of being in state $s$ at time $t+u$ given the current state $r$ at time $u$, depends only on the elapsed time $t$. Transition matrix  $\mathbf{P}(t)$ contains these probabilities such that the rows sum up to 1.  The hazards are defined by
\[
q_{rs}=\lim_{\Delta t \to 0}\frac{P\big(Y(t+\Delta t)=s|Y(t)=r\big)}{\Delta t},
\]
for $r\neq s$. The matrix with off-diagonal entries $q_{rs}$ and diagonal entries $q_{rr}= -\sum _{r\neq s} q_{rs}$ is the  generator matrix $\mathbf{Q}$. Given $\mathbf{Q}$, the solution for $\mathbf{P}(t)$ subject to $\mathbf{P}(0)= \mathbf{I}$ is $\mathbf{P}(t)=\exp(t \mathbf{Q} )$, see, e.g., Cox \cite{cox2017theory}. 

Time-dependent models can be defined by using proportional hazards model for  transition $r$ to $s$, $r\neq s$ as follows
\begin{eqnarray}
q_{rs}(t)=q_{rs.0}(t)\exp\big(\boldsymbol{\beta}_{rs}^{\top}\mathbf{x}\big),
\label{eq:hazardmodel}
\end{eqnarray}
where $q_{rs.0}(t)$ is the baseline hazard function, $\mathbf{x}$ is a covariate vector and $\boldsymbol{\beta}_{rs}^{\top}$ is vector of unknown parameters. We focus on the nonparametric estimation of $q_{rs.0}(t)$ with splines. Each hazard can be approximated by the exponential of a linear combination of $K_{rs}$ spline base functions $B_{k}(t)$ and regression coefficients $\alpha_{rs.k} \in \mathbb{R}$ as follows
\begin{equation}
q_{rs.0}(t)= \exp\left(\sum_{k=1}^{K_{rs}}\alpha_{rs.k}B_{k}(t)\right). \label{haz.1}
\end{equation}

Let the number of spline basis be large (usually $K_{rs} \geq 10$) and define the vector of coefficients by $\bs{\alpha}_{rs} = (\alpha_{rs.1}, \ldots, \alpha_{rs.K_{rs}})^{\top}$ for $r\neq s$. To control the wiggliness of the estimated hazard, each $\bs{\alpha} _{rs}$ is associated to a quadratic penalty $\lambda_{rs}\bs{\alpha}_{rs}^{\top}\mathbf{S}_{rs}\bs{\alpha}_{rs}$, which is employed in estimation. The smoothing parameter $\lambda_{rs}$ controls the trade-off between model fit and model smoothness. Large values for the smoothing parameters, $\lambda_{rs} \rightarrow \infty $, lead to a log-linear estimate of $q_{rs.0}$, while $\lambda _{rs} = 0$ results in an unpenalised regression spline estimate \citep{wood2006generalized}. For the spline base functions, $B_{k}(t)$, we use cubic regression splines which have convenient mathematical properties for multistate modelling. However, the method is implemented in a way that is easy to employ other spline definitions and corresponding penalties.


\subsection{Likelihood function}
\noindent Given a multistate model, maximum likelihood inference can be used to analyse longitudinal data. For interval-censored transition times, the likelihood function is constructed using transition probabilities. Let the state space be $\mathcal{S}=\{1,2,..,D\}$, with $D$ the dead state.

Let  $Y_1,...,Y_{n}$ be a  series of states observed at times $t_1,...,t_n$, respectively. The inference is conditional on the first observed state. For $Y_2,...,Y_{n}$, the distribution is
\begin{eqnarray}
\prob\left(Y_{n}=y_{n},...,Y_{2}=y_{2}|Y_{1}=y_{1}, \bs{\theta},  \mbf{t}, \mbf{X}\right),\label{eq:jointdistr}
\end{eqnarray}
where ${\bs{\theta}}$ is the vector with the model parameters, $\mbf{t}=(t_1,...,t_n)^\top$, and the $n\times p$ matrix $\mbf{X}$ contains the values of the $p$ covariates at each of the $n$ time points. A conditional Markov assumption is used to define the distribution (\ref{eq:jointdistr}) as
\begin{eqnarray*}
\prod_{j=2}^{n}\prob\left(Y_j=y_j|Y_{j-1}=y_{j-1},\bs{\theta}, t_{j-1}, \mbf{x}_{j-1}\right),
\end{eqnarray*}
where $ \mbf{x}_{j-1}$ is the $(j-1)^{th}$ row in $\mbf{X}$. Given $N$ individuals, the likelihood function is given by
\begin{eqnarray}
L(\bs{\theta}) = \prod_{i=1}^{N}\prod_{j=2}^{n_i}\ \prob\left(Y_{ij}=y_{ij}|Y_{ij-1}=y_{ij-1}\right),
\label{eq:likelihood}
\end{eqnarray}
where $n_i$ is the number of observation times for individual $i$. 

If time of death is known, the likelihood contribution of the interval $(t_{n-1}, t_n]$ in which an individual is observed alive at time $t_{n-1}$ and subsequently dead at time $t_{n}$ is given by $\sum_{s=1}^{D-1}P\left(Y_{n}=s|Y_{n-1}=y_{n-1}\right)q_{sD}(t_{n-1})$. A similar definition of the likelihood can be found in Jackson \cite{jackson2011multi}.

\subsection{Piecewise-constant hazards}\label{pwca}
\noindent Time-dependency of the hazard model (\ref{eq:hazardmodel}) can be taken into account by using a piecewise-constant approximation. In longitudinal data for continuous-time models, follow-up times often vary across individuals. If that is the case, the individual-specific follow-up times can be used to define the piecewise-constant approximation for the individual likelihood contributions. This implies that a transition probability such $P\left(Y_{j}=y_{j}|Y_{j-1}=y_{j-1}\right)$ is derived by using $\mathbf{Q}(t_{j-1})$ to estimate $\mathbf{P}(t_{j-1},t_j)$ by $\exp ((t_j - t_{j-1})\mathbf{Q}(t_{j-1}))$. It is also possible to impose a fixed grid to the piecewise-constant approximation as described in \cite{van2008}. For most applications, both methods lead to similar result and the method described in this Section is preferable as it is less computationally extensive \cite{van2014multi}. 

\section{Penalised maximum likelihood estimation}\label{main_method}
\subsection{Penalised log-likelihood function}
\noindent For each hazard, let the number of splines basis dimension be large enough to allow for flexible modelling. Define the full set of parameter by $\bs{\theta} $ and the penalty matrix by $\mathbf{S}_{\bs{\lambda}}$. This is a block diagonal matrix with blocks $\lambda_{rs} \mathbf{S}_{rs}$ for penalising splines parameters of transition $r$ to $s$ and zeros elsewhere. The amount of smoothing is controlled by adding a smoothness penalty to the log-likelihood function. Let $\ell (\bs{\theta})$ be the logarithm of the likelihood function. The penalised log-likelihood function is 
\begin{eqnarray}
\ell_p(\mbox{\boldmath$\theta$}) &=& \ell (\mbox{\boldmath$\theta$}) - \frac{1}{2} \mbox{\boldmath$\theta$}^\top \mathbf{S}_{\mbox{\boldmath$\lambda$}} \mbox{\boldmath$\theta$}. \label{eq:penlike}
\end{eqnarray}

\subsection{Parameter estimation}\label{par_est}
\noindent Given a piecewise-constant approximation to the time-dependency in the hazard model (\ref{eq:hazardmodel}), a scoring algorithm can be used to maximise the penalised log-likelihood function (\ref{eq:penlike}) \cite{machado2017flexible}. For a given multistate model, if more than one hazard is specified with splines, then estimation of $\bs{\lambda}$ by direct grid search can be computationally burdensome. 

There are methods available for automatic smoothing parameters estimation within the penalised likelihood framework; see Wood \cite{wood2006generalized} and Radice et al. \cite{radice2016copula}. For their method, the derivatives of the penalised log-likelihood function have to be split into the derivatives with relation to the linear predictors, and the derivatives of the linear predictor with relation to the model parameters. The direct use of their methods in multistate models leads to large sparse matrices that are difficult to deal with. 

Marra et al. \cite{marra2016simultaneous} developed a more general method for automatic smoothing, which uses the gradient and the Hessian (or Fisher information matrix) as a whole instead of components that make them up.  The method consist of two parts. First, given a value for the smoothing parameters, we aim to find an estimate of the model parameters. Second, we use such an estimate to find an update for the smoothing parameters. We next describe how to perform the first part of the method.

Let $\mathbf{g}_{p}^{[a]} = \mathbf{g}^{[a]} - \mathbf{S}_{\bs{\lambda}}\bs{\theta}^{[a]}$ and $\bm{\mathcal{I}}_p^{[a]} = \bm{\mathcal{I}}^{[a]} + \mathbf{S}_{\bs{\lambda}}$ represent the penalised gradient and negative of the penalised hessian matrix at iteration $a$, respectively, where $\mathbf{g}^{[a]} = \partial \ell (\mbox{\boldmath$\theta$}) / \partial \mbox{\boldmath$\theta$}|_{\mbox{\boldmath$\theta$} =  \mbox{\boldmath$\theta$}^{[a]} }$ and $\bm{ \mathcal{I} }^{[a]} = - \partial ^2 \ell(\mbox{\boldmath$\theta$}) / \partial \mbox{\boldmath$\theta$} \partial \mbox{\boldmath$\theta$}^{\top}|_{\mbox{\boldmath$\theta$} =  \mbox{\boldmath$\theta$}^{[a]} }$. For fixed value of  $\widehat{\bs{\lambda}}$, the $a^{th}$ estimate of $\bs{\theta}$ can be updated by 
\begin{equation}
\bs{\theta}^{[a+1]} = \left(\bm{\mathcal{I}}^{[a]} + \mathbf{S}_{\widehat{\bs{\lambda}}}\right)^{-1}\sqrt{\bm{\mathcal{I}}^{[a]}} \mathbf{z}^{[a]},
\end{equation}
where $\mathbf{z}^{[a]} = \sqrt{\bm{\mathcal{I}}^{[a]}} \bs{\theta}^{[a]} + \bs{\epsilon}^{[a]}$ and $\bm{\epsilon}^{[a]} = \sqrt{\bm{\mathcal{I}}^{[a]}}^{-1}\mathbf{g}^{[a]}$. 

This parametrisation of the model-parameters estimator allows for a well founded formulation of the smoothing parameters selection presented in Section \ref{sp_est} \cite{marra2016simultaneous}. See \ref{ap_derive} for a justification for this parametrisation of the parameters estimator. Calculating the second derivatives of the probability matrix can be intractable; see Kalbfleish and Lawless \cite{kalbfleisch1985}. We use an approximation to the Fisher information matrix that involves only the first order derivatives of the penalised log-likelihood function; see \ref{ap_scoring}.

\subsection{Smoothing parameters estimation}\label{sp_est}
\noindent The penalised maximum likelihood approach described in Section \ref{par_est} can only estimate model parameters, $\bs{\theta}$, for fixed vector of smoothing parameters, $\bs{\lambda}$. In this section, we briefly discuss the automatic smoothing parameters estimation presented in Marra et al. \cite{marra2016simultaneous}. 

From likelihood theory, $\bm{\epsilon} \sim \mathcal{N}(0, \mathbf{I})$ and $\mathbf{z} \sim \mathcal{N}(\bm{\mu}_{\mathbf{z}}, \mathbf{I})$, where $\mathbf{I}$ is the identity matrix, $\bm{\mu}_{\mathbf{z}}= \sqrt{\bm{\mathcal{I}}}\bs{\theta}$ and $\bs{\theta}$ is the true parameter vector. The predicted value vector for $\mathbf{z}$ is $\widehat{\bs{\mu}}_{\mathbf{z}} = \sqrt{\bm{\mathcal{I}}}\widehat{\bs{\theta}}= \mathbf{A}_{\widehat{{\bs{\lambda}}}}\mathbf{z}$, where $\mathbf{A}_{\widehat{\bs{\lambda}}} = \sqrt{\bm{\mathcal{I}}}(\bm{\mathcal{I}} + \mathbf{S}_{\widehat{\bs{\lambda}}})^{-1}\sqrt{\bm{\mathcal{I}}}$. The smoothing parameter vector is estimated to minimise
\begin{eqnarray*}
\mathbb{E}(||\bs{\mu}_{\mbf{z}} - \widehat{\bs{\mu}_{\mbf{z}}} ||^2) = \mathbb{E}(||\mbf{z} - \mbf{A}_{\widehat{\bs{\lambda}}}\mbf{z} ||^2) - c + 2tr(\mbf{A}_{\widehat{\bs{\lambda}}}),
\end{eqnarray*}
where $c$ is a constant. In practice, $\bs{\lambda}$ is estimated by minimising the Un-Biased Risk Estimator (UBRE; Craven and Wahba, 1979)
\begin{eqnarray}
\mathcal{V}(\bs{\lambda}) = ||\mbf{z} - \mbf{A}_{\bs{\lambda}}\mbf{z} ||^2 - c + 2tr(\mbf{A}_{\bs{\lambda}}). \label{eq.:ubre}
\end{eqnarray}
Equation (\ref{eq.:ubre}) can be minimised using the automatic smoothing parameters selection method developed by Wood (2004) or in principle by using a general-purpose optimiser.

\subsection{Summary of the algorithm}\label{alg_summary}
\noindent The methods described in Sections \ref{par_est} and \ref{sp_est} can be used to define an algorithm that iterates until the parameter estimator satisfies $\max |\bs{\theta}^{[a+1]} - \bs{\theta}^{[a]}| < \delta$ for a suitable small positive value \cite{radice2016copula}. The two steps of the algorithm are as follow:
\begin{description}
\item[\textbf{Step 1:}] For fixed smoothing parameters $\bs{\lambda}^{[a]}$, find an estimate of $\bs{\theta}$:
\begin{equation}
\bs{\theta}^{[a+1]} =  \underset{\bs{\theta}}{\operatorname{argmax}} \ \ell_p(\mbox{\boldmath$\theta$}). \nonumber
\end{equation}  
\item[\textbf{Step 2:}] Given the estimate $\bs{\theta}^{[a+1]}$, find an estimate of $\bs{\lambda}$ using equation (\ref{eq.:ubre}):
\begin{equation}
\bs{\lambda}^{[a+1]} = \underset{\bs{\lambda}}{\operatorname{argmin}} \ \mathcal{V}(\bs{\lambda}). \nonumber
\end{equation}

\end{description}

\subsection{Confidence intervals}
\noindent The distribution of the penalised maximum likelihood estimator can be used to construct confidence intervals for the estimate $\widehat{\bs{\theta}}$ and functions of them, such as the hazards and probability matrix \cite{wood2006generalized}. Let $\mathbf{V}_{\bs{\theta}}$ represent the covariance matrix of $\widehat{\bs{\theta}}$ at convergence. From large sample theory, samples of the estimate $\widehat{\bs{\theta}}$ can be drawn from $ N(\widehat{\bs{\theta}}, \mathbf{V}_{\bs{\theta}}).$ Confidence intervals for functions of the model parameters can be constructed as follows:
\begin{description}
\item[\textbf{Step 1:}] Draw $n$ vectors from $N(\widehat{\bs{\theta}}, \mathbf{V}_{\bs{\theta}}).$
\item[\textbf{Step 2:}] Calculate the value of the function of interest at each simulated value.
\item[\textbf{Step 3:}] Using the simulated values of the function, calculate the lower ($\varsigma/2$) and upper ($1-\varsigma$), quantiles. 
\end{description} 
The parameter $\varsigma$ is usually set to 0.05. In this paper, we approximate the covariance matrix $\mathbf{V}_{\bs{\theta}}$ by the inverse of the matrix $\mathbf{M}$ described in \ref{ap_derive}.

\section{Simulation study}
\noindent We perform a small simulation study to analyse the performance of the method presented in Section \ref{main_method} for modelling time-dependency in multistate processes. The simulation is described for an illness-death model with a log-normal distribution with parameters $\mu=1.25$ and $\sigma = 1$ for transition 1 to 2, an exponential distribution with rate $\exp(-2.5)$ for transition 1 to 3 and a Gompertz distribution with rate $\exp(-2.5)$ and shape $0.1$ for 2 to 3. 

Let $T_{rs} = T_{rs|u}$ represent the time to the event $s$ conditional on being in state $r$ at time $u>0$. If state at $u$ is 1, then the time of transition to the next state can be obtained by taking $T = \min\{T_{12}, T_{13}\}$. If $T=T_{12}$ then, the next state is 2, otherwise the next state is 3. If state is 2, then the time of the next state is $T_{23}$. The event times $T_{12}$ and $T_{13}$ are simulated using the functions \textsf{rgengamma()} and \textsf{rgompertz()}, respectively, in \textsf{R} \cite{jackson2016flexsurv}. The transition times $T_{23}$ can be simulated by sampling from uniform distribution and using the inversion method. 

We perform $R=100$ simulation replications. The sample size is $N=200$ individuals. The time scale is years since baseline, i.e., time since the beginning of the study. The length of follow-up is one year and the length of the study is 15 years. This leads to interval-censored transition times for transitions 1 to 2 and known time of transitions into the dead state. 

The package \textsf{mgcv} \cite{wood2007mgcv} in R is used to set the design and penalty matrices. The number of knots for each hazard is $K=10$, hence the model has a total of 30 parameters. We use cubic regression splines, in which case the knots are placed using the percentiles of the observation times. Therefore, knots placement is different for every sample. The multistate model with splines is then estimated using the procedure described in Section \ref{main_method}. The smoothing parameters are estimated using the general-purpose optimiser \textsf{optim} in \textsf{R}.  
\begin{figure}[t!]\centering
\includegraphics[scale=0.55 ]{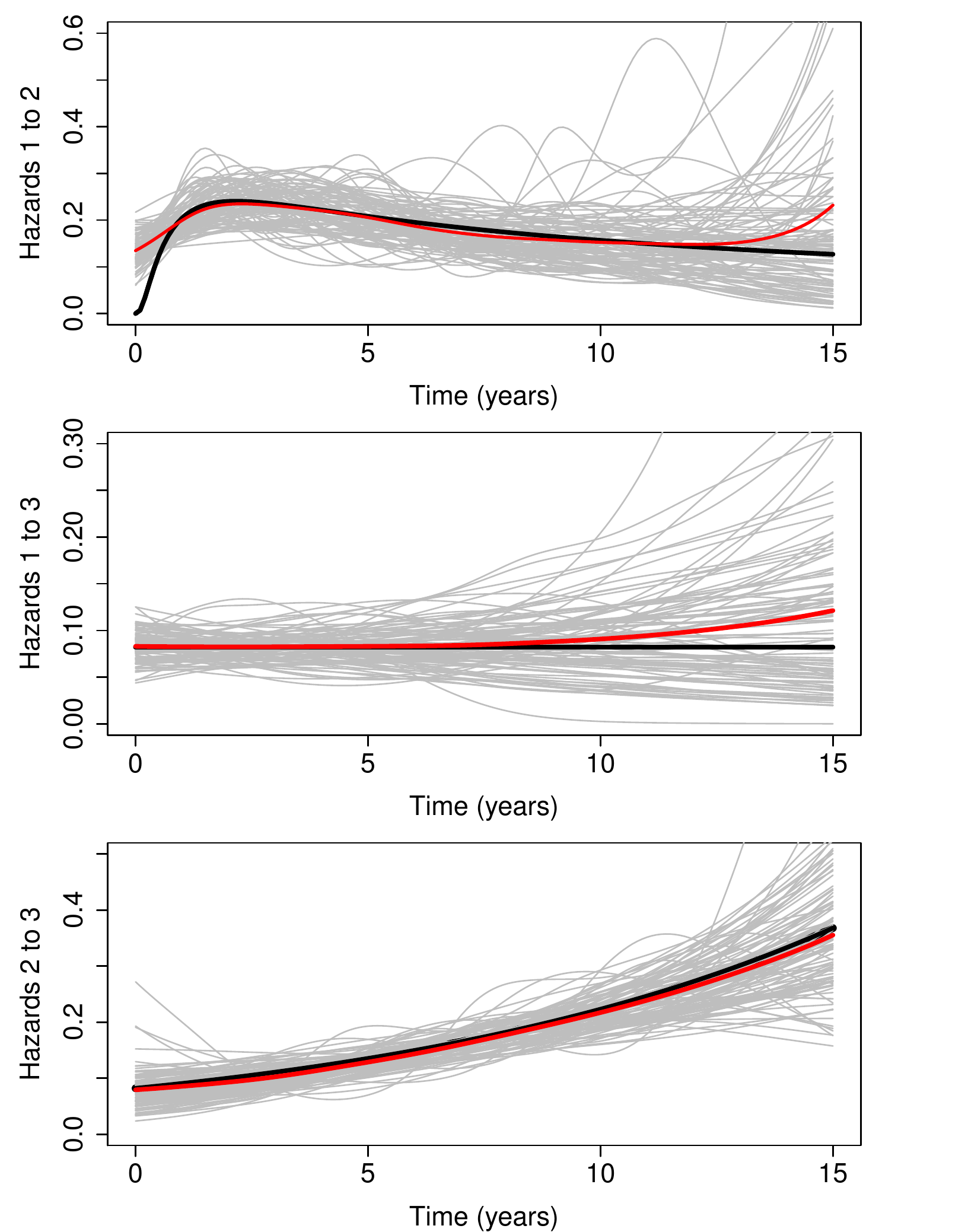}
\caption{\label{sim_plot} Simulation study: true (black lines), estimated (grey lines) and mean estimated (red lines) hazards for the illness-death model for 100 replications }
\end{figure}

Figure \ref{sim_plot} illustrates the true (black lines), estimated (grey lines) and mean estimated (red lines) hazards for the illness-death model for 100 replications. Some estimated hazards seem to over or under estimate the true hazards; however, the mean estimated hazards are very close to the true hazard curves. The discrepancy between the hazards towards the end of study is due to scarcity of data after 10 years. In particular, the method satisfactorily estimates the nonlinear trend underlying the hazard for transition 1 to 2. 

Table \ref{sim_table} presents the results of the simulation in terms of transition probabilities. It shows the ten-year transition probabilities for the true model, the mean estimated ten-year transition probabilities and the bias. The results show that the multistate model with splines can estimate well transition probabilities for the ten-year time interval $(0,10]$.

The findings from the simulation results are twofold. First, they indicate that the proposed method is able to estimate nonlinear, log-linear and linear hazards in the presence of interval censoring. Second, they show that the piecewise-constant approximation to the transition probabilities provides satisfactory results, as we are able to recover the true curves and ten-year transition probabilities. 
\begin{table}[t!]
\caption{Simulation study to investigate the performance of the multistate models with splines for modelling time-dependent processes. Mean and bias for $R=100$ replications. Absolute bias less than $x$ is denoted by $\lceil x \rceil$ }
\centering
\begin{tabular*}{\columnwidth}{@{}l@{\extracolsep{\fill}}l@{\extracolsep{\fill}}c@{\extracolsep{\fill}}r@{\extracolsep{\fill}}r@{\extracolsep{\fill}}r@{\extracolsep{\fill}}l@{\extracolsep{\fill}}c@{\extracolsep{\fill}}c@{\extracolsep{\fill}}c@{}}
 \hline
Transition probabilities & \multicolumn{1}{r}{True} & \multicolumn{1}{l}{Mean} & \multicolumn{1}{l}{Bias} \\ \hline
 $p_{11}(0,10)$ &0.065 &  0.065 & $\lceil 0.001\rceil$ \\ 
  $p_{12}(0,10)$ &0.232 & 0.239 & 0.008 \\
 $p_{13}(0,10)$ &0.703 & 0.695 & -0.008 \\
 $p_{22}(0,10)$ &0.246 & 0.263 & 0.018 \\
 $p_{23}(0,10)$ &0.754 & 0.737& -0.018 \\ \hline
\end{tabular*}
\label{sim_table}
\end{table}

\section{Application to CAV data}
\noindent We fit a progressive three-state model for the CAV data defined as in Figure \ref{fig1}. Because time of death is known within one day, rather than being interval censored, the likelihood contribution of individuals observed in state $r<3$ at time $t$ and dead at time $t^*>t$ are given by $\sum_{s=1}^{2}P\left(Y(t^*)=s|Y(t)= r \right)q_{s3}(t)$. As described in Section \ref{pwca}, transition probabilities for the likelihood function are calculated by using a piecewise-constant approximation to the hazards. For the CAV data, the mean length of follow-up times is $1.622$ years with standard deviation of  $0.972$ and median $1.258$. Assuming that change of health status can be assessed in intervals of approximately $1.2$ years, we can use the data to define the grid for the piecewise-constant approximation. 

\begin{figure}[t!]\centering
\includegraphics[scale=0.5 ]{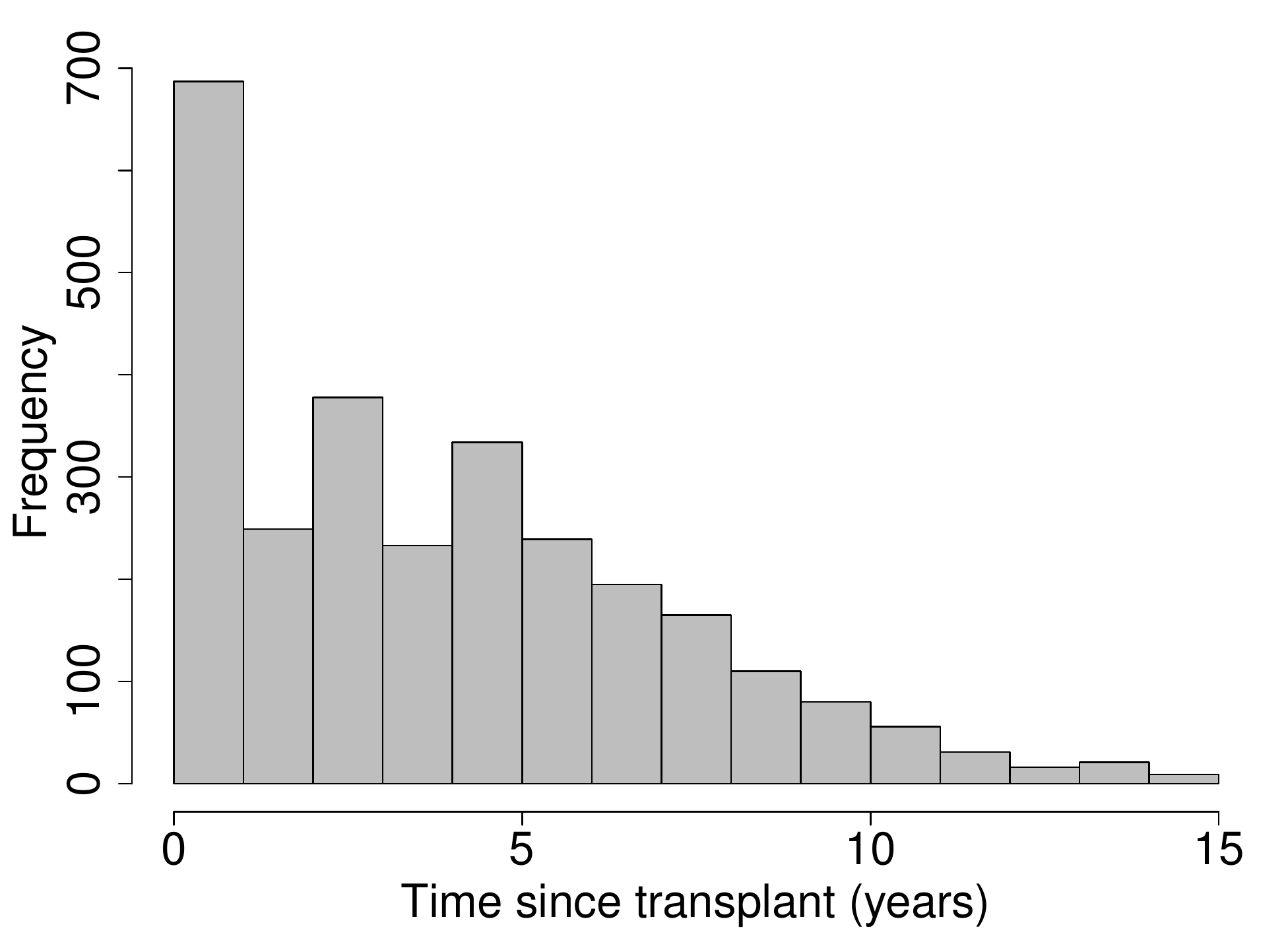}
\caption{\label{fig3} Histogram of time since transplant in the CAV data}
\end{figure}

Let $t$ represent time since baseline. The proportional hazard model with splines is specified with dependence on donor age ($dage$) and primary diagnosis of ischaemic heart disease ($IHD$):
\begin{equation}
q_{rs}(t)= \exp\left(\sum_{k=1}^{10}\alpha_{rs.k}B_{k}(t)  + \beta_1dage + \beta_2IHD \right), \label{haz_cav}
\end{equation} 
where $(r,s)\in\{ (1,2), (1,3), (2,3) \}$ and $B_{k}(t)$ are known spline basis function. We use penalised cubic regression splines. The knots are placed considering the percentiles of the observation times. This is a key factor for fitting multistate models with splines. Because multistate data can become scarce close to the end of study, there might not be enough information to estimate some basis coefficients. Fitting multistate models with $P$-splines \cite{eilers1996} might not be possible for some applications as it requires the knots to be equally spaced. In this case, some knots can be placed where there is no data. Figure \ref{fig3} illustrates the histogram of time since transplant for the CAV data. 

For the analysis to follow, the design and penalty matrices are set up using the package \textsf{mgcv} in \textsf{R}. As indicated in (\ref{haz_cav}), the hazards are modelled with 10 knots each, hence the total number of parameters is 32. The vector of smoothing parameters is $\bm{\lambda}^\top = (\lambda_{12}, \lambda_{13}, \lambda_{23})$. The multistate model with splines is then estimated using the procedure described in Section \ref{main_method}. The smoothing parameters are estimated using the general-purpose optimiser \textsf{optim} in \textsf{R}.  

\begin{figure}[t!]\centering
\includegraphics[scale=0.7 ]{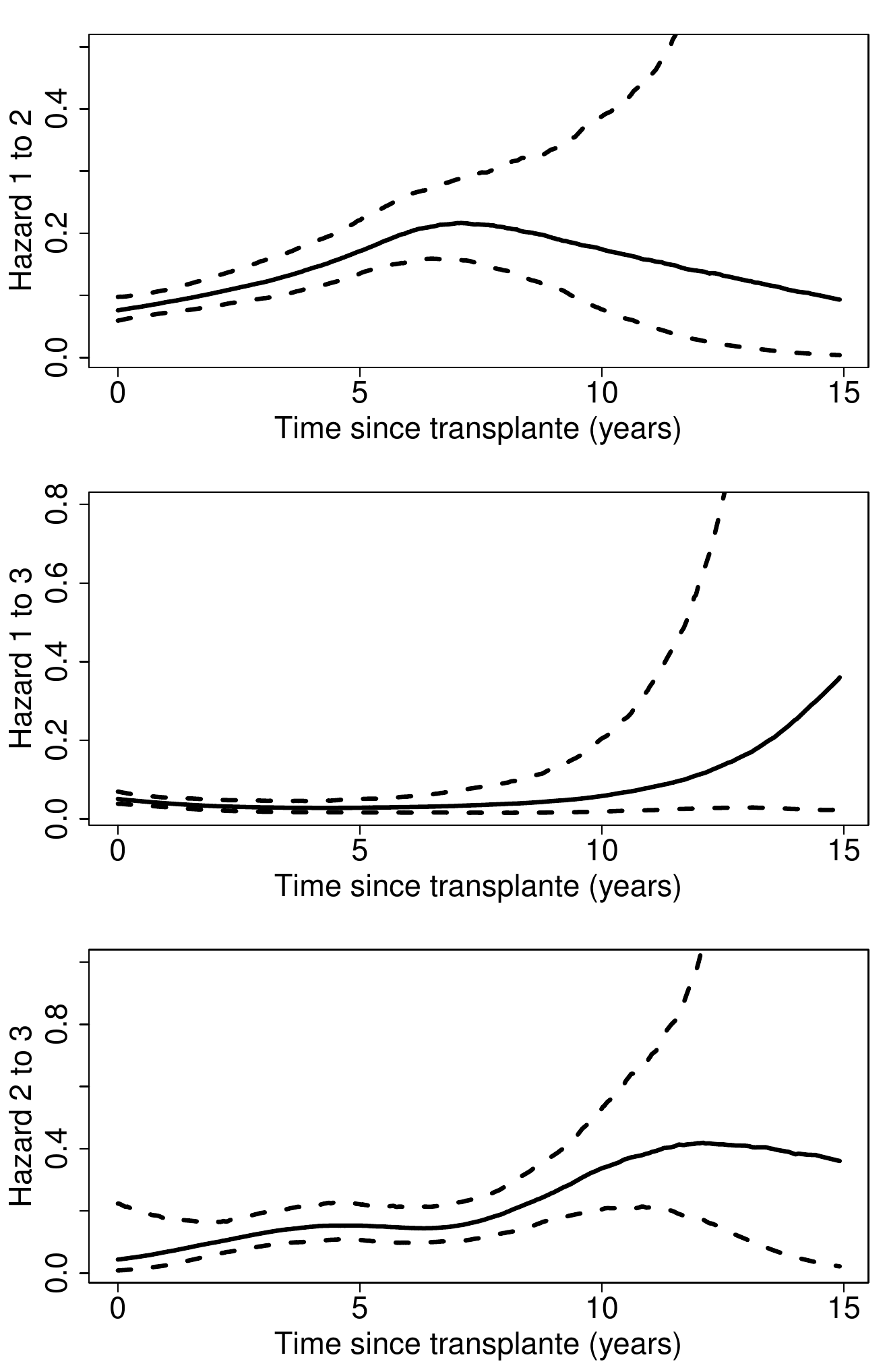}
\caption{\label{fig2} Estimated smooth hazards for subjects with IHD and with donor age of 26 (solid lines), with 95\% confidence intervals (dashed lines)}
\end{figure}

The estimated smooth hazards for subjects with $IHD$ and donor age of 26 (solid lines) and  $95\%$ confidence intervals (dashed lines) are presented in Figure~\ref{fig2}. The risk of moving from state 1 (healthy) to state 2 (CAV) increases until approximately 8 years after transplant, but decreases afterwards. The risk of going from state 1 to state 3 (dead) is very low and almost constant until approximately 10 years since transplant, but increases pretty steep afterwards. The transition intensity from state 2 to state 3 is quite volatile and upwards until 10 years after transplant and decreasing afterwards. The confidence intervals are fairly wide after approximately 10 years. That is because data become scarce after 10 years. For the parametric part of the model, $\widehat{\beta_1} = 0.018$ and $\widehat{\beta_2} = 0.274$ indicating that donor age and $IHD$ increase the risks of disease progression and death. The vector of smoothing parameters is estimated at $\widehat{\bs{\lambda}} = (47.145, 41.668, 10.716)^\top$.

Although estimated hazards gives insightful information about the risks of moving across states, interpretation is more straightforward when transition probabilities are considered. For subject with $IHD$ and with donor age of 26, the five-year transition probabilities  are estimated at
\begin{eqnarray*}
\widehat{\mbf{P}}(0,5)=
{\small
\left(
\begin{array}{lll}
  0.475\ (0.412,0.529) &0.291\ (0.250, 0.335) &0.234\ (0.197, 0.286)\\
  0 &0.579\ (0.428, 0.675) &0.421\ (0.325, 0.572) \\
  0& 0& 1
\end{array}
\right),
}\label{predict_matrix}
\end{eqnarray*}
with 95\% confidence interval (in brackets) obtained using $1000$ simulations. A transition probability can be interpreted as follows. A subject with $IHD$ and donor age of 26 has a 29\% chance of being in the CAV five years later. 

\section{Discussion}\label{s:discuss}
\noindent This paper presents a practical framework for estimating multistate models with splines for interval-censored data. The new estimation method is made possible by rewriting the optimisation problem using a penalised general likelihood estimation \cite{marra2016simultaneous}.

The method is applied to an illness-death model without recovery. We aim to illustrate the feasibility of the method and its usage for flexible time-dependent modelling. There should not be a problem to apply the method for more complex multistate processes with backwards transitions, as long as there are enough observations for transitions modelled with splines. 

The small simulation study and application show the importance of this method for flexible modelling of time-dependent processes. We show in a simulation that the method can recover nonlinear, log-liner and linear hazards. Furthermore, we illustrate with an application that the method can give insightful information on the functional form underlying the hazards. 

The automatic smoothing parameters estimation as described in Marra et al. \citep{marra2016simultaneous} requires the Hessian or the Fisher information for estimation. We show through a simulation and an application that an approximation to the Fisher information matrix, which only uses the first order derivatives of the log-likelihood, performs well on estimation. This is relevant for interval-censored data as calculating the second derivatives of the transition probabilities can be intractable.

As discussed in Titman \citep{titman2011}, CAV is a progressive disease even though backwards transition are recorded, due to measurement errors. The work presented here can be extended to allow for misclassification of states \citep{jackson2003multistate}. This poses extra difficulty for estimation as derivative free algorithms, e.g., a quasi-Newton algorithm is required to maximise the penalised log-likelihood function.

The \textsf{msm} package \citep{jackson2011multi} is designed to model time-homogeneous multistate models. However, it is possible to fit some time-dependent models, such as Gompertz and splines (without penalties) models. In this case, time-dependency is also approached by using a piecewise-constant approximation to the hazards. Therefore, this research can also be seen as a generalisation of the \textsf{msm} package, which allows for flexible modelling of the time-dependency.


\section*{Acknowledgements}
This research was supported by CNPq - Brazil [249308/2013-4].
\vspace*{-8pt}

\section*{References}

\bibliographystyle{elsarticle-num}
\bibliography{mybibfile}

\appendix

\section{Justification for the parameter estimators}\label{ap_scoring}
\noindent For easy reference, we derive the parametrisation of the model-parameter estimators  as in Marra et. al \cite{marra2016simultaneous}. A first-order Taylor expansion of $\mathbf{g}_p^{[a+1]}$ about the current fit $\bs{\theta}^{[a]}$ is given by
\begin{equation}
\mathbf{g}_p^{[a+1]} \approx \mathbf{g}_p^{[a]} + \bm{\mathcal{H}}_p^{[a]}(\bs{\theta}^{[a+1]} - \bs{\theta}^{[a]}),\label{grad_update}
\end{equation}
where $\mathbf{g}_p^{[a+1]} = \mathbf{g}^{[a]} - \mathbf{S}_{\widehat{\bs{\lambda}}}\bs{\theta}^{[a]}$ and $ \bm{\mathcal{H}}_p^{[a]} = \bm{\mathcal{H}}^{[a]} - \mathbf{S}_{\widehat{\bs{\lambda}}}$. Let us define $\bm{\mathcal{I}}^{[a]} = - \bm{\mathcal{H}}^{[a]}$. A new fit  $\bs{\theta}^{[a+1]}$ is obtained by taking the right-hand side of equation (\ref{grad_update}) to be zero
\begin{eqnarray*}
\mathbf{0} &=& \mathbf{g}_p^{[a]} + \left(-\bm{\mathcal{I}}^{[a]} - \mathbf{S}_{\widehat{\bs{\lambda}}}\right)(\bs{\theta}^{[a+1]} - \bs{\theta}^{[a]})\\
\mathbf{g}_p^{[a]} &=&  \left(\bm{\mathcal{I}}^{[a]} + \mathbf{S}_{\widehat{\bs{\lambda}}}\right)(\bs{\theta}^{[a+1]} - \bs{\theta}^{[a]})\\
\mathbf{g}^{[a]} - \mathbf{S}_{\widehat{\bs{\lambda}}}\bs{\theta}^{[a]} &=& \left(\bm{\mathcal{I}}^{[a]} + \mathbf{S}_{\widehat{\bs{\lambda}}}\right)\bs{\theta}^{[a+1]} - \bm{\mathcal{I}}^{[a]}\bs{\theta}^{[a]}  - \mathbf{S}_{\widehat{\bs{\lambda}}}\bs{\theta}^{[a]} \\
\left(\bm{\mathcal{I}}^{[a]} + \mathbf{S}_{\widehat{\bs{\lambda}}}\right)\bs{\theta}^{[a+1]} &=& \mathbf{g}^{[a]} + \bm{\mathcal{I}}^{[a]}\bs{\theta}^{[a]}\\
\bs{\theta}^{[a+1]} &=& \left(\bm{\mathcal{I}}^{[a]} + \mathbf{S}_{\widehat{\bs{\lambda}}}\right)^{-1} \sqrt{\bm{\mathcal{I}}^{[a]}}\left(\sqrt{\bm{\mathcal{I}}^{[a]}}\bs{\theta}^{[a]} + \sqrt{\bm{\mathcal{I}}^{[a]}}^{-1}\mathbf{g}^{[a]} \right). 
\end{eqnarray*}
Therefore, the new fit for the parameter estimator can be expressed as 
\begin{equation}
\bs{\theta}^{[a+1]} = \left(\bm{\mathcal{I}}^{[a]} + \mathbf{S}_{\widehat{\bs{\lambda}}}\right)^{-1} \sqrt{\bm{\mathcal{I}}^{[a]}}\mathbf{z}^{[a]},
\end{equation}
where $\mathbf{z}^{[a]} = \sqrt{\bm{\mathcal{I}}^{[a]}} \bs{\theta}^{[a]} + \bs{\epsilon}^{[a]}$ with $\bm{\epsilon}^{[a]} = \sqrt{\bm{\mathcal{I}}^{[a]}}^{-1}\mathbf{g}^{[a]}$.

\section{Derivatives}\label{ap_derive}
\noindent In this appendix, we derive the gradient vector and an approximation to the Fisher information matrix. The description to follow is also presented in Van den Hout \cite{van2014multi}. 

Given piecewise-constant intensities, the likelihood contribution for an observed time interval $(t_1,t_2]$ is defined using a constant generator matrix $\mbf{Q}=\mbf{Q}(t_1)$. For the eigenvalues of $\mbf{Q}$ given by
$\mbf{b}=(b_1,...,b_D)$, define $\mbf{B}=\mbox{diag}(\mbf{b})$. Given matrix $\mbf{A}$ with the eigenvectors as columns, the eigenvalue decomposition is $\mbf{Q}=\mbf{A}\mbf{B}\mbf{A}^{-1}$.  The transition probability matrix $\mbf{P}(t)=\mbf{P}(t_1,t_2)$ for elapsed time $t=t_2-t_1$ is given by
\[
\mbf{P}(t)=\mbf{A}\ \mbox{diag}\left(e^{b_1t},...,e^{b_Dt}\right)\ \mbf{A}^{-1}.
\]

As described in Kalbfleisch and Lawless \cite{kalbfleisch1985}, the derivative of $\mbf{P}(t)$ can be obtained as
\[
\frac{\partial}{\partial \theta_k}\mbf{P}(t)=\mbf{A}
\mbf{V}_k
\mbf{A}^{-1},
\]
where $\mbf{V}_k$ is the $D \times D $ matrix with $(l,m)$ entry
\begin{eqnarray*}
\left\{
\begin{array}{ll}
 g_{lm}^{(k)}\left[\exp(b_lt)-\exp(b_mt)\right]/(b_l-b_m) & l\neq m\\\\
 g_{ll}^{(k)}t\exp(b_lt)  & l= m,
\end{array}
\right.
\end{eqnarray*}
where $g_{lm}^{(k)}$ is the $(l,m)$ entry in $\mbf{G}^{(k)}=\mbf{A}\partial \mbf{Q}/\partial \theta_k\mbf{A}^{-1}$.

Let  $\mbf{g}(\bs{\theta})$ denote the $q\times 1$ gradient vector. The $k$th entry of $\mbf{g}(\bs{\theta})$ is given by
\[
\sum_{i=1}^N\sum_{j=2}^{n_i}\frac{\partial}{\partial \theta_k}\log P(Y_{ij} = y_{ij} | Y_{ij-1} = y_{ij-1} )\,.
\]
The Fisher information matrix is given by
$\bs{\mathcal{I}}(\bs{\theta})=\E\left[\mbf{g}(\bs{\theta})\mbf{g}(\bs{\theta})^\top \right]$, which can be estimated by
defining the $q\times q$ matrix $\mbf{M}({\bs{\theta}})$ with $(k,l)$ entry
\[
\sum_{i=1}^N\sum_{j=2}^{n_i}\frac{\partial}{\partial \theta_k}\log P(Y_{ij} = y_{ij} | Y_{ij-1} = y_{ij-1} )
\frac{\partial}{\partial \theta_l}\log P(Y_{ij} = y_{ij} | Y_{ij-1} = y_{ij-1} )\,.
\]

\end{document}